\documentclass[aps,prb,twocolumn,showpacs]{revtex4}
\usepackage{graphicx}

\begin{document}

\title{Zero-temperature transition and correlation-length exponent of the
frustrated XY model on a honeycomb lattice}

\author{Enzo Granato}

\address{Laborat\'orio Associado de Sensores e Materiais,
Instituto Nacional de Pesquisas Espaciais,  12227-010 S\~ao Jos\'e
dos Campos, SP Brazil}

\begin{abstract}
Phase coherence and vortex order in the fully frustrated XY model on
a two-dimensional honeycomb lattice are studied by extensive Monte
Carlo simulations using the parallel tempering method and
finite-size scaling. No evidence is found for an equilibrium
order-disorder or a spin/vortex-glass transition, suggested in
previous simulation works. Instead, the scaling analysis of
correlations of phase and vortex variables in the full equilibrated
system is consistent with a phase transition where the critical
temperature vanishes and the correlation lengths diverge as a
power-law with decreasing temperatures and corresponding critical
exponents $\nu_{ph}$ and $\nu_{v}$. This behavior and the  near
agreement of the critical exponents suggest a zero-temperature
transition scenario where phase and vortex variables remain coupled
on large length scales.

\end{abstract}


\maketitle

\section{Introduction}

Two-dimensional frustrated XY models \cite{villain,teitel}, in which
vortices form a dense incommensurate lattice, have attracted
considerable interest as a possible two-dimensional vortex glass
without quenched disorder \cite{halsey,teitelf,reid,eg07,eg08}  or a
structural glass of supercooled liquids \cite{llkim,tarjus}. In
superconducting systems with pinning described by XY models,
frustration can be introduced by applying an external magnetic field
as in periodic arrays of Josephson junctions
\cite{carini,zant,baek}, superconducting wire networks
\cite{yu,ling,xiao} and superconducting thin films with a periodic
pattern of nanoholes \cite{valles}. The frustration parameter $f$
sets the average density of vortices in the lattice of pinning
centers and can be tuned by varying the strength of the external
field \cite{teitel}. Depending on the structure of the lattice of
pinning centers and the value of $f$, a commensurate vortex lattice
is favored in the ground state, which leads to discrete symmetries
in addition to the continuous symmetry of the phase variables
characterizing the superconducting order parameter. In this case,
the phase transitions and resistive behavior of the system are
reasonably well understood for simple low-order commensurate phases
such as $f=1/2$, on a square lattice \cite{teitel}, and $f=1/3$, on
a triangular lattice \cite{shih85} of pinning centers. However, when
the vortex lattice is incommensurate with the pinning centers, both
the nature of the equilibrium phase transition and of the
low-temperature state in the thermodynamic limit is less clear. This
is the case of Josephson-junction arrays \cite{carini,baek} and
superconducting wire networks \cite{yu,ling} on a square lattice,
described by the frustrated XY model with irrational $f$, which has
been extensively studied by various methods
\cite{halsey,teitelf,eg07,eg08,llkim,tang,park}. Another interesting
physical realization but much less investigated so far, should occur
in Josephson-junction arrays and superconducting wire networks
\cite{xiao} on a honeycomb lattice, and superconducting films with a
triangular pattern of nanoholes \cite{valles}, which can be
described by the fully frustrated XY model ($f=1/2$) on a honeycomb
lattice.

In early Monte Carlo (MC) simulations of the fully frustrated XY
model on a honeycomb lattice \cite{shih85}, a phase transition at a
nonzero temperature in the Koterlitz-Thouless universality class was
suggested,  based on the saturation of the specific-heat peak with
increasing system sizes and the apparent jump in the helicity
modulus. On the other hand, calculations of the the spin-glass order
parameter by MC simulations \cite{reid} suggested that, instead of
an order-disorder transition, a spin-glass transition should take
place approximately at the same temperature. In contrast, from MC
simulations of a similar model in the vortex representation
\cite{leeteitel}, it was argued that only a crossover region rather
than an equilibrium phase transition should occur at any nonzero
temperature, since the energy of domain-wall excitations which
disorder the ground state was found to approach a finite constant
for increasing system sizes. However, the free energy of domain-wall
excitations in the frustrated XY model obtained analytically in the
phase representation \cite{korshu} was found to increase linearly
with the system length, but with an extremely small numerical
coefficient. As a consequence, although vortex ordering could be
possible in the thermodynamic limit, domain-wall excitations would
only be negligible for system sizes which are much beyond the ones
studied numerically or even experimentally. It is unclear if the
behavior observed in the earlier numerical works \cite{shih85,reid}
could be a signature of such vortex ordering or the effect of slow
dynamics which prevents to observe the true equilibrium behavior.
Given these conflicting results, it should be of interest to further
investigate the fully frustrated XY model using a MC method which
can insure full equilibration and a scaling analysis of both phase
and vortex correlations.

In this work we study phase coherence and vortex order in the fully
frustrated XY model on a honeycomb lattice, using the
parallel-tempering method (exchange MC method)
\cite{nemoto,marinari} to obtain equilibrium configurations of the
system. This method has been shown to reduce significantly the long
equilibration times in glassy systems \cite{nemoto,cooper}. To study
the equilibrium phase transitions we use numerical data in the
temperature regime in which full equilibration can be insured and
employ a scaling analysis of the correlation lengths of the phase
and vortex variables to extrapolate to the low-temperature and
large-system limits. No evidence of an equilibrium order-disorder
phase transition or even a spin/vortex-glass transition is found at
nonzero temperatures. Our results, however, are consistent with an
equilibrium zero-temperature transition, where the critical
temperature vanishes ($T_c=0$) and the correlation lengths diverge
as a power-law with decreasing temperatures and corresponding
critical exponents $\nu_{ph}$ and $\nu_{v}$. This transition has
important consequences for the resistivity behavior and
current-voltage scaling \cite{fisher,eg98,eg07} at low temperatures
in the superconducting systems described by the frustrated XY model.
Moreover, the near agreement of the critical exponents estimated
numerically suggests a $T_c=0$ transition scenario, where phase and
vortex variables remain coupled on large length scales. This is in
contrast with the  $T_c=0$ transition in the frustrated XY model on
a square lattice at irrational frustration, where a decoupled
scenario was found recently \cite{eg08}.

\section{Model and Monte Carlo simulation}

We consider a uniformly frustrated XY model described by the
Hamiltonian \cite{teitel}
\begin{equation}
H=-J\sum_{<ij>}\cos(\theta_i -\theta_j-A_{ij}) ,  \label{model}
\end{equation}
where $\theta_i$ is a phase variable defined at the sites $i$ of a
two-dimensional honeycomb lattice (Fig. 1), representing the local
angle of the XY spin with an arbitrary fixed direction. $J>0$ is a
uniform ferromagnetic coupling and  $A_{ij}$ is a gauge-invariant
quantity constrained to be $\sum_{ij}A_{ij} = 2 \pi f$ around each
plaquette of the lattice. The parameter $f$ controls the frustration
of the system. For the fully frustrated case considered here,
$f=1/2$. In the calculations we choose a gauge where $A_{ij} = 2 \pi
f n_i/2$ on the bonds along the  horizontal rows numbered by the
integer $n_i$ and $A_{ij}=0$ on the vertical bonds of the lattice.

As a model of an array of superconducting grains coupled by
Josephson junctions, the phase $\theta_i$ in Eq. (\ref{model})
corresponds to the phase of the  superconducting order parameter of
the grains, $J$ is the Josephson coupling between grains and
$A_{ij}$ is the line integral of the vector potential $\vec A$ due
to an external magnetic field $\vec B=\nabla \times \vec A$ applied
perpendicular to the array. The magnetic flux in each plaquette in
units of the flux quantum $\Phi_o=hc/2e$ can be written as $2 \pi f$
with the frustration parameter $f$ corresponding to the number of
flux quantum per plaquette.

In the Monte Carlo simulations we use the parallel tempering method
\cite{nemoto} to obtain equilibrium configurations. In this method,
many replicas of the system with different temperatures are
simulated simultaneously and the corresponding configurations are
allowed to be exchanged with a probability satisfying detailed
balance. The exchange process allows the configurations of the
system to explore the temperature space, being cooled down and
warmed up, and the system can escape more easily from metastable
minima at low temperatures. Full equilibration can be insured within
reasonable computer time in systems of sufficiently small sizes
\cite{nemoto}. Without the replica exchange step, the method reduces
to conventional MC simulations at different temperatures. We
performed MC simulations using the heat-bath algorithm for each
replica, simultaneously and independently, for a few MC passes.
Periodic boundary conditions were used on lattices with rhombic
geometry (Fig. 1) of side $L$, containing $N=2 L^2$ sites. Exchange
of pairs of replica configurations at temperatures $T_i$ and $T_j$
and energies $E_i$ and $E_j$ is attempted with probability
$min(1,\exp(-\Delta))$, where $\Delta = (1/T_i - 1/T_j)(E_j -E_i)$,
using the Metropolis scheme. The equilibration time to reach thermal
equilibrium can be measured as the average number of MC passes
required for each replica to travel over the whole temperature
range. We used typically $4 \times 10^6$ MC passes for equilibration
with up to $120$ replicas and $1.6 \times 10^7$ MC passes for
calculations of average quantities.

\begin{figure}
\includegraphics[bb= -0.5cm -0cm  9cm   5cm, width=7.5cm]{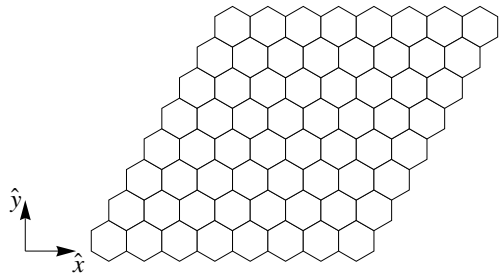}
\caption{ Honeycomb lattice and the coordinate axes. }
\end{figure}

\section{Correlation length and Scaling analysis}

For the study of phase coherence, we consider the overlap order
parameter \cite{bhatt} of the phase variables defined as
$q_{ph}(j)=\exp(i\theta_j^1-i\theta_j^2)$, where 1 and 2 denote two
thermally independent copies of the system, with the same parameters
$J$ and $f$ in the model of Eq. \ref{model}. At high temperatures,
where each copy is thermally disordered, the correlation function
\begin{equation}
C_{ph}(r)=\frac{1}{N}\sum_j<q_{ph}(j)q_{ph}(j+r)>
\end{equation}
is short ranged, decaying exponentially with $r$, while at low
temperatures it is long ranged if an  ordered phase or a glassy
phase exits. The latter possibility was suggested in ref.
\onlinecite{reid} based on MC simulations and the glass phase was
characterized by the behavior of the Edwards-Anderson order
parameter $q_{EA}$, which corresponds here to $<\sum_j
q_{ph}(j)/N>$.

The correlation length in the finite system $\xi_{ph}(L,T)$ can be
obtained as \cite{cooper,cooper2}
\begin{equation}
\xi_{ph}(L,T)=\frac{1}{2\sin(k_o/2)}(\frac{S_{ph}(0)}{S_{ph}(k_o)}-1)^{1/2}
, \label{xieq}
\end{equation}
where $S_{ph}(k)$ is the Fourier transform of $C_{ph}(r)$  and $k_o$
is the smallest nonzero wave vector in the finite system. This
expression for the correlation length $\xi(L,T)$, which is both
temperature and size dependent, can be obtained from the correlation
length $\xi(T)$ in the large system limit, $\xi(T)^2=
-\frac{1}{S(k)}\frac{\partial S(k)}{\partial k^2} \vert_{k=0}$, as a
finite-difference approximation taking also into account the lattice
periodicity \cite{cooper}. Thus, $\xi(L,T)$ tends to $\xi(T)$ for
large $L$, when correlations are short ranged. If there is
long-range order, $\xi(L,T) \sim L^{1+d/2} $, while if there is only
algebraic order $ \xi(L,T) \sim L$. Note that the correlation length
defined in Eq. (\ref{xieq}), in terms of the overlap order parameter
$q_{ph}(j)$, may have a different magnitude from the correlation
length defined in terms of a single copy of the system
$\exp(i\theta_J^1)$. However, they should have the same leading
scaling behavior near the transition.

Analogous expressions are used to determine the correlation length
for vortex variables $\xi_{v}(L,T)$,
\begin{equation}
\xi_{v}(L,T)=\frac{1}{2\sin(k_o/2)}(\frac{S_{v}(0)}{S_{v}(k_o)}-1)^{1/2}
,
\end{equation}
in terms of the overlap order parameter $q_v(p)=v_p^1 v_p^2$ of the
net vorticity $v_p=n_p-f$. The vorticity $n_p$ is defined as
$n_p=\sum_{ij}(\theta_i-\theta_j-A_{ij})/2\pi$, where the summation
is taken around the elementary plaquette located at sites $p$ of the
dual lattice, formed by plaquette centers, and the gauge-invariant
phase difference is restricted to the interval $[-\pi,\pi]$. For the
fully frustrated case, $f=1/2$, the vortex variables $v_p= \pm 1/2$
are Ising-like variables and equivalent to the chirality variables
originally introduced by Villain \cite{villain}.

Finte-size scaling can be used to extrapolate the behavior of the
system to the large-system limit and temperatures near the
transition. In the scaling analysis of the correlation length
\cite{cooper}, we consider the dimensionless ratio $\xi(L,T)/L$
which, for a continuous transition, should satisfy the scaling form
\begin{equation}
\xi(L,T)/L=G((T-T_c)L^{1/\nu}) \label{scalxi}
\end{equation}
where $T_c$ is the critical temperature and $\nu$ is the critical
exponent of the power-law divergent correlation length
$\xi(T)\propto |T-T_c|^{-\nu}$. The scaling function $G(x)$ has the
properties: $G(0)=C$, a constant, and $G(x)\rightarrow x^{-\nu}$ as
$x\rightarrow \infty $. This scaling form implies that data for the
scaled correlation length $\xi(L,T)/L$ as a function of temperature,
for different system sizes $L$, should come together for decreasing
temperatures and cross at the same temperature $T=T_c$. In addition,
the data should splay out for different system sizes with slopes
determined by the critical exponent $\nu$. If the data satisfy the
scaling form of Eq. (\ref{scalxi}) then we can infer that the
correlation length diverges as  a power-law  $\xi(T)\propto
|T-T_c|^{-\nu}$ and estimate the critical exponent $\nu$. A very
large value of $\nu$ would be an indication that the correlation
length may diverge exponentially, $\xi(T)\propto e^{c/|T-T_c|^{-\nu
^\prime}}$, rather than as a power law. In particular, for the
standard (unfrustrated) XY model, which is critical at temperatures
equal and below $T_c$, curves of $\xi(L,T)/L$ as function of
decreasing temperatures for different system sizes  merge
\cite{cooper} near $T_c$ and remain $L$-independent for $T < T_c$. .

\section{Results and Discussion}

We study the behavior of the correlations of phase and vortex
variables {\it in thermal equilibrium}, which implies that we should
only use the numerical data obtained from the MC simulations in the
temperature range where full equilibration is achieved. To check
equilibration, we follow the trajectory in the temperature space of
a replica starting at the highest temperature, where the system
equilibrates fast even without the replica exchange process
\cite{eg08}. In equilibrium, this replica should be able to explore
the whole temperature range. For the system sizes studied, $L \ge 24
$, equilibration was only possible for temperatures above $T_f
\approx 0.11$, despite the improvement of the parallel tempering
method. Below $T_f$, the configurations of the replicas at different
temperatures cannot be warmed up and cooled down. Thus $T_f$ can be
regarded as a freezing temperature, below which the system remains
trapped in metastable configurations within the available time scale
of the present simulation. It is interesting to note that $T_f$
agrees with the apparent glass transition temperature $T_g = 0.11$
observed in earlier MC simulations \cite{reid}. However, as will be
shown in the following, $T_f$ does not correspond to the critical
temperature of an equilibrium glass transition. Instead, it should
be regarded as the characteristic temperature of a dynamical
freezing transition.

\begin{figure}
\includegraphics[bb= 0.5cm -0cm  10cm   7cm, width=7.5cm]{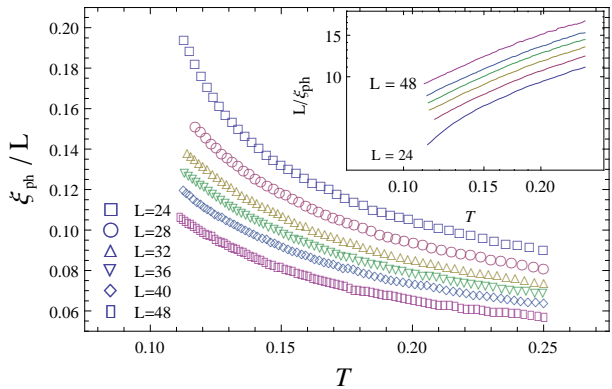}
\caption{ Scaled correlation length of phase variables $\xi_{ph}/L$
for different system sizes $L$. Inset: Log-log plot of $L/\xi_{ph}$
{\it versus} $T$ for the corresponding system sizes  $L$.  }
\end{figure}

Figures 2 and 3 show the temperature dependence of the scaled
correlation length in the $\hat{x}$-direction for phase variables,
$\xi_{ph}/L$, and for vortex variables, $\xi_{v}/L$, in the
temperature range ($T \ge T_f)$ where full equilibration was
possible and for different system sizes. Both quantities increase
faster on lowering the temperature as the system size $L$ increases
indicating a divergent length scale for decreasing temperatures.
However, for fixed temperature they decrease with $L$ even at the
lowest available temperature. Therefore, the curves for the
different system sizes do not cross or merge at a common
temperature, as would be expected from the scaling form of Eq.
(\ref{scalxi}), if a transition occurs in this temperature range.
This is more evident in the insets of Figs. 2 and 3, where lines
joining the data for different system sizes are presented as a
log-log plot of $L/\xi(L,T)$ {\it versus } $T$.

\begin{figure}
\includegraphics[bb= 0.5cm -0cm  10cm   7cm, width=7.5cm]{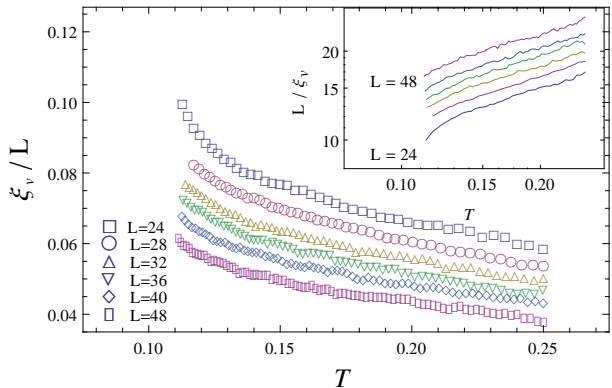}
\caption{ Scaled correlation length of vortex variables $\xi_{v}/L$
for different system sizes $L$. Inset: Log-log plot of $L/\xi_{v}$
{\it versus} $T$ for the corresponding system sizes  $L$.  }
\end{figure}

For the sake of comparison to the behavior of the vortex correlation
for the fully frustrated case ($f=1/2$) in Figs. 3, we show the
results of additional calculations  when the frustration parameter
is $f=1/3$, in Fig. 4. In this case, a hexagonal vortex pattern
commensurate with the honeycomb dual lattice is possible in the
ground state and an order-disorder phase transition is expected at
finite temperature from symmetry arguments. From the earlier work
\cite{shih85}, this transition takes place at a critical temperature
$T_c\sim 0.23$. Indeed, the curves in Fig. 4 cross for different
system sizes at a common temperature $T_c =0.226(1)$, which is
compatible with this estimate.

\begin{figure}
\includegraphics[bb= 2cm 3cm  19cm   16cm, width=7.5 cm]{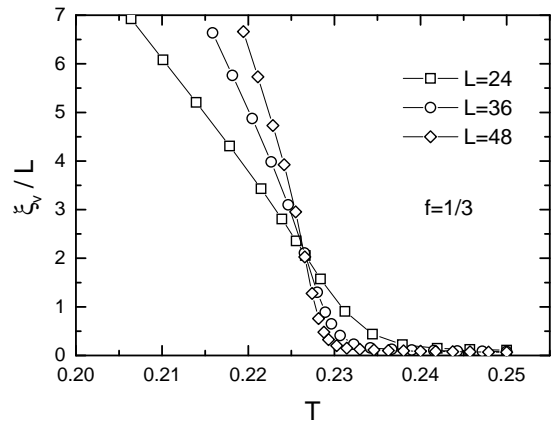}
\caption{ Scaled correlation length of vortex variables $\xi_{v}/L$
for different system sizes $L$, when the frustration parameter is
$f=1/3$. }
\end{figure}

For the fully frustrated case, $f=1/2$, the lack of intersection of
the curves of  $\xi_{ph}/L$ and  $ \xi_{v}/L$ for different system
sizes at a common temperature in Figs. 2 and 3, suggests that phase
coherence and vortex order, or even glass-like order,  can only
occur at $T << T_f$, which is not accessible in our calculations, or
else only at $T=0$. The latter scenario corresponds to a phase
transition where $T_c=0$ and the correlation length $\xi(T)$ is
finite at any nonzero $T$ but diverges when approaching $T=0$. To
verify the possibility of such zero-temperature transition, we first
consider the scaling of the total correlation function $\chi$
defined as $\chi_{ph}=\sum_r C_{ph}(r)$ and $\chi_v=\sum_r C_{v}(r)$
for phase and vortex variables, respectively. In the absence of
finite-size effects, $\chi$ should diverge  as \cite{jain,kawa87}
\begin{equation}
\chi = \frac{A}{(T-T_c)^\gamma} + B,  \label{sucfit}
\end{equation}
where $A$ and $B$ are constants and  $\gamma$ is a critical
exponent. $B$ represents possible background corrections to scaling.
Using data for a large system size, where finite-size effects are
negligible, we can then determine $T_c$ from the best numerical fit
according to this scaling form. Using a least-square fit
\cite{press} we have obtained the error estimate $\sigma(T_c)$ when
fitting $Log(\chi-B)$ against $Log(T-T_c)$ to a straight line,
assuming different values of $T_c$.  The error estimate is defined
as
\begin{equation}
\sigma^2 = \frac{1}{N_p}\sum_{i=1}^{N_p} (y_i - f(x_i))^2,
\label{error}
\end{equation}
where $N_p$ is the number of data points $\{ x_i , y_i \}$ and
$f(x_i)$ are the values of the fitting function at the corresponding
data points. Figs. 5 and 6 show the dependence of the error $\sigma$
on $T_c$, for phase and vortex variables, respectively, for $L=40$.
In both cases, $T_c=0$ gives the lowest error, which suggests that a
zero-temperature transition is possible. The insets in the Figures
show the corresponding best fits, which provide estimates of the
critical exponents $\gamma_{ph}=1.57(1)$ and $\gamma_v=1.14(1)$.

\begin{figure}
\includegraphics[bb= 2cm 3cm  19cm   16cm, width=7.5 cm]{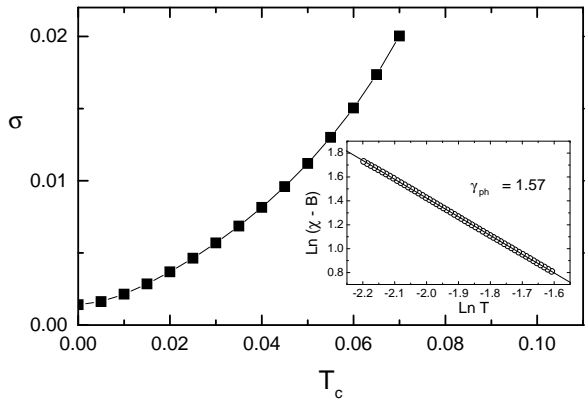}
\caption{ Variation of the fitting error  $\sigma$ against $T_c$,
for the total phase correlation $\chi_{ph}$ according to Eq.
\ref{sucfit}, with $B=2.6$. Inset: log-log fit assuming $T_c=0$. }
\end{figure}

\begin{figure}
\includegraphics[bb= 2cm 3cm  19cm   16cm, width=7.5 cm]{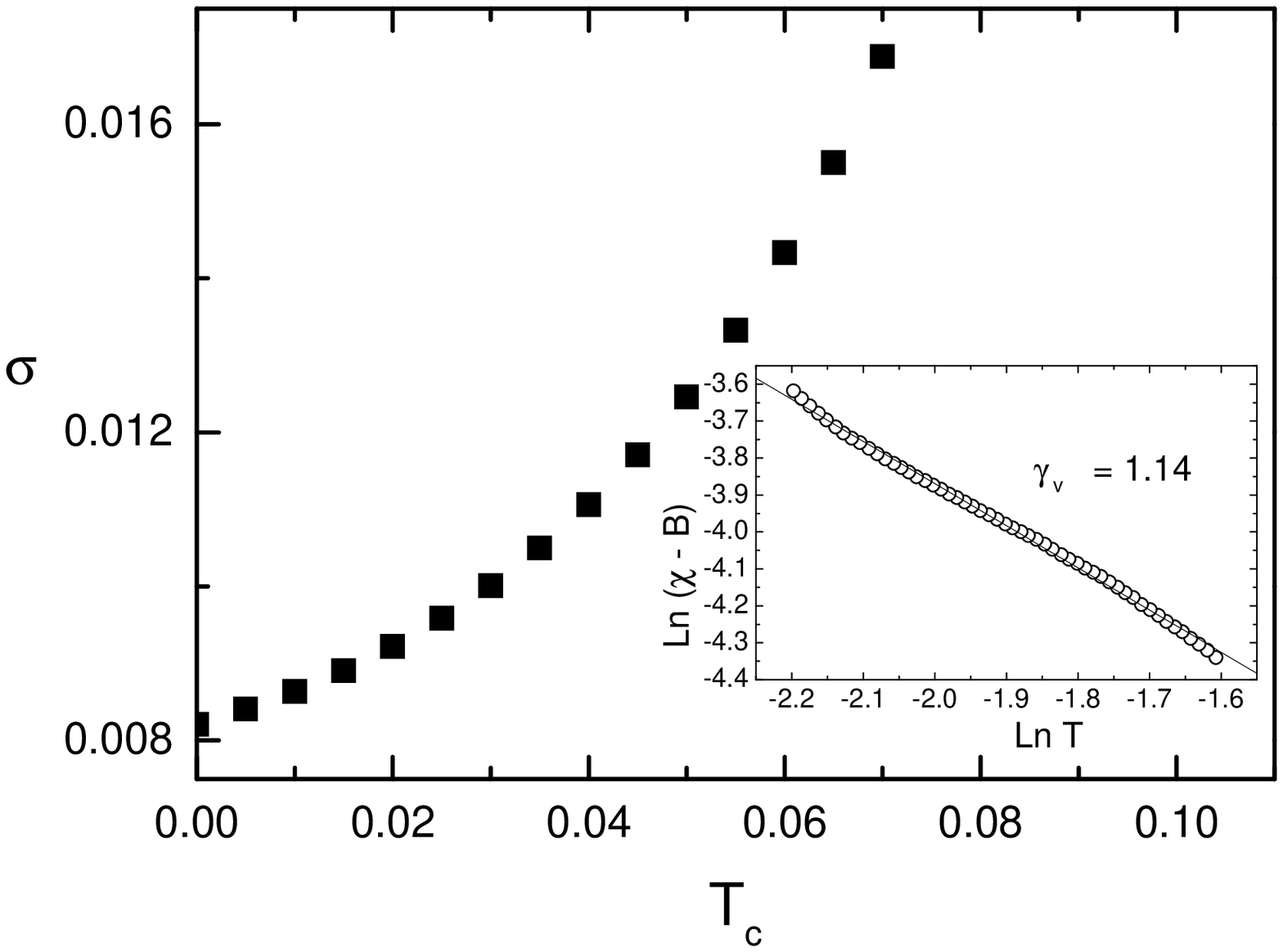}
\caption{ Variation of the fitting error  $\sigma$ against $T_c$,
for the total vortex correlation $\chi_{v}$ according to Eq.
\ref{sucfit}, with $B=0.04$. Inset: log-log fit assuming $T_c=0$. }
\end{figure}

The $T_c=0$ scenario can be further verified from the finite-size
scaling analysis of the correlation length $\xi(L,T)$. In this case,
the data for $\xi(L,T)$ should satisfy the finite-size scaling form
of Eq. (\ref{scalxi}) with $T_c=0$. The best data collapse is
obtained by adjusting a single parameter, $\nu$, providing an
estimate of this critical exponent. We use the following procedure,
which is a simplified version of more general methods of measuring
data collapse \cite{night,seno}. The standard finite-size scaling
expansion near $T_c$,
\begin{equation}
G(x) = a_o + a_1 x + a_2 x^2 + \dots, \label{fit}
\end{equation}
truncated to low order ($ O(x^5))$, is used as a smooth
interpolation function, which is convenient both for numerical data
fitting and for providing a measure of the data collapse as the
error defined by Eq. (\ref{error}). Then we determine $\nu$ from the
best least-square fit of $\xi(L,T)/L$ to this function with $x=T
L^{1/\nu}$. Data for the largest temperatures are successively
removed from the data collapse if this leads to a decrease of the
fitting error, since in this case such data are presumably outside
the asymptotic scaling regime where Eq. (\ref{scalxi}) applies. The
best data collapse is obtained  for temperatures $ T < 0.16$.  Fig.
7 shows that indeed the data for the phase variables $\xi_{ph}$
satisfy the scaling form with an exponent $\nu_{ph}=1.29(15)$.
Similarly, for the vortex variables, we obtain from the dada
collapse in Fig. 8 the critical exponent $\nu_{v}=1.03(25)$,
although the data collapse is not as good. These exponents agree
with each other within the estimated errors. It should be noted,
however, that the scaling behavior in Eq. (\ref{scalxi}) has only
been verified here in a limited range of temperatures since there is
no available data below $T_f$, due to the lack of equilibration.
This can be seen more clearly  by reploting the data collapse as
$L/\xi(L,T)$ {\it versus} $y \equiv L T^\nu$, as presented in the
insets of Figs. 7 and 8. Indeed, $L/\xi(L,T)$ tends to a linear
behavior for large $y$ as expected but the finite limit for $y
\rightarrow 0$ can not be explicitly verified.

\begin{figure}
\includegraphics[bb= 0.5cm -0cm  10cm   7cm, width=7.5cm]{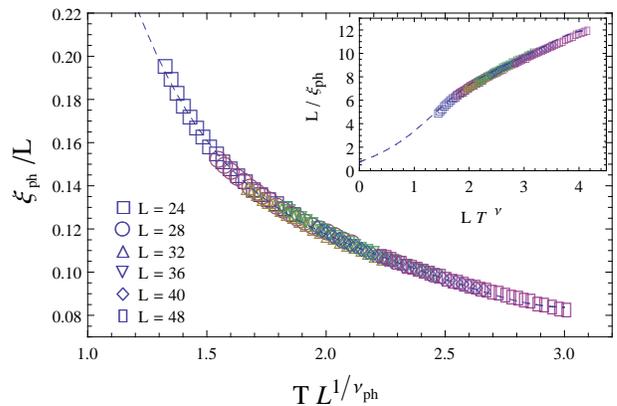}
\caption{ Scaling plot according to Eq. (\ref{scalxi}), assuming
$T_c=0$, for the phase correlation length $\xi_{ph}$, giving the
estimate $\nu_{ph}=1.29(15)$. Dashed line is a fit to the data with
Eq. (\ref{fit}). Inset: Replot of data collapse in terms of the
variable $ L T^{\nu_{ph}}$. }
\end{figure}

The zero-temperature transition, which was inferred here from the
behavior of correlations at finite temperatures as discussed above,
is also consistent with the ground-state properties of the model. As
first pointed out by Shih and Stroud \cite{shih85}, the ground state
is highly degenerate with an infinite number of different vortex
configurations. Therefore, correlations at zero temperature,
obtained by averaging over the different configurations, can decay
to zero for large distances. One possibility is that such decay
obeys a power law. The triangular antiferromagnetic Ising model
\cite{stephenson}, for example, display such behavior. Another
possibility is that correlations decay exponentially as in the
antiferromagnetic Ising model on a Kagom\'e lattice \cite{huse}. The
$T_c=0$ scenario for the present model, favors a power-law decay of
vortex correlations. In fact, the correlation function exponent
$\eta _v$ associated with the transition can be obtained from the
scaling law $\gamma=\nu(2-\eta )$ and the above estimates of
$\gamma_v$ and $\nu_v$, giving $\eta_v=0.90$, which is greater than
zero implying a decay of the correlation as $C_v(r)\propto
r^{-\eta_v}$, at zero temperature. To verify this behavior more
quantitatively, we have also performed additional calculations of
the finite-size dependence of the total correlation functions
$\chi_{ph}$ and $\chi_v$ near $T=0$. However, because full
equilibration was only possible for temperatures above the dynamical
freezing transition $T_f$ with the present Monte Carlo method, we
have to rely on the approximation of using low-energy states close
to the ground state to perform the configuration averages. The
low-energy states were obtained using the simulated-annealing method
\cite{press} by gradual annealing from initial temperatures $T
\gtrsim T_f$. Fig. 9 shows a log-log plot of the finite-size
behavior of $\chi_{ph}$ and $\chi_v$ obtained by averaging over
$800$ low-energy sates within a range of $0.16 \%$ above the known
ground-state energy \cite{shih85} $E_g=-1.2071$. If correlations
decay as a power-law in the ground state, then the finite-size
behavior $\chi \propto L^{2-\eta}$ is expected near $T=0$, for
sufficiently large systems. This behavior is consistent with the
data in Fig. 9 and from a log-log fit  we estimate $\eta_{ph}=0.4(2)
$ and $\eta_{v}=0.4(3) $, which are compatible within the estimated
errors  with the corresponding values obtained from $\gamma$ and
$\nu$, using the finite-temperature scaling analysis.

\begin{figure}
\includegraphics[bb= 0.5cm -0cm  10cm   7cm, width=7.5cm]{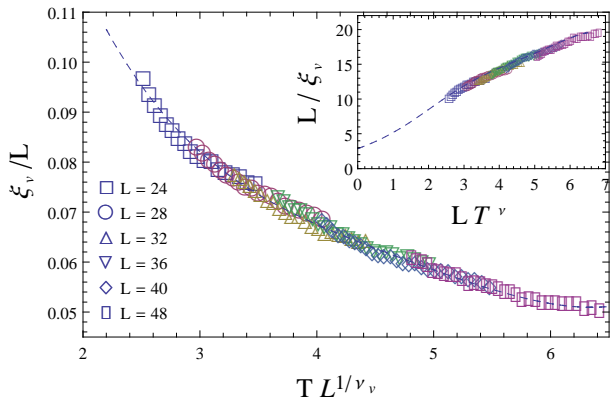}
\caption{ Scaling plot according to Eq. (\ref{scalxi}), assuming
$T_c=0$, for the vortex correlation length $\xi_{v}$, giving the
estimate $\nu_v=1.03(25)$. Dashed line is a fit to the data with Eq.
(\ref{fit}). Inset: Replot of the data collapse in terms of the
variable $ L T^{\nu_v}$.}
\end{figure}

A particular feature of the frustrated XY model on the honeycomb
lattice is the structure of  the dual lattice, where the chiralities
are defined, which has the form of a triangular lattice. Since the
chiralities are Ising-like variables with antiferromagnetic
interactions, if interactions further than nearest-neighbors are
neglected, there is a geometric frustration effect which may appear
similar to the case of the triangular antiferromagnetic Ising model.
The exact solution of this model \cite{stephenson} shows that there
is no phase transition at nonzero temperatures ( $T_c=0$), which is
the same scenario we find in the present case. However, while the
correlation length for this Ising model diverges exponentially
\cite{egunp}, $ \xi \propto e^{2/T}$, corresponding to $\nu
\rightarrow \infty $ in Eq. (\ref{scalxi}), in the present case we
find $\nu_v \sim 1$, suggesting a power-law correlation  and
correspondingly different ground state properties.

The near agreement of our estimates of the critical exponents for
vortex variables $\nu_v$ and phase variables $\nu_{ph}$, assuming
power-law divergent correlations, suggests that the $T_c=0$ critical
behavior may be described by a single divergent length scale.  In a
zero-temperature transition scenario where there is no decoupling,
phase and vortex variables remain coupled on large length scales and
one expects that the corresponding correlation lengths should
diverge with a common critical exponent $\nu_{ph}=\nu_v=\nu$. It is
interesting to note that this behavior is in contrast with that
found for the two-dimensional XY model at irrational frustration on
a square lattice \cite{eg08}, where a decoupled zero-temperature
transition \cite{kawa} was found with significant different
exponents $\nu_v >> \nu_{ph}$.

\begin{figure}
\includegraphics[bb= 1cm 3.5cm  19cm   16cm, width=7.5 cm]{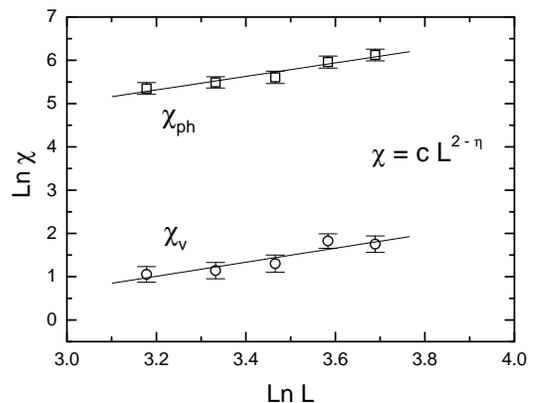}
\caption{ Finite-size behavior of the total correlation function of
phase  $\chi_{ph}$ and vortex $\chi_{v}$ variables obtained from
low-energy states, using simulated annealing. Lines correspond to
log-log fits, which give the estimates $\eta_{ph}=0.4(2) $ and
$\eta_{v}= 0.4(3)$}
\end{figure}

An additional important feature of the zero-temperature transition
found from the above analysis, is the temperature dependence of  the
relaxation time for phase and vortex equilibrium fluctuations. If
$T_c=0$ then the divergence of the relaxation time $\tau$ as $T
\rightarrow T_c$ is determined by thermal activation \cite{fisher}.
Thus, we expect that $\tau$ should increase exponentially with the
inverse of temperature, corresponding to a dynamical exponent $z
\rightarrow \infty$ in the usual power-law assumption $\tau\propto
|T-T_c|^{-z \nu}$. To verify this behavior, we have in addition
calculated the relaxation time $\tau$ for different temperatures
from the autocorrelation function of phase and vortex fluctuations,
$C_{ph}(t)$ and $C_{v}(t)$,  as
\begin{equation}
 \tau_{ph,v} =\frac{1}{C_{ph,v}(0)}\int_0^\infty  dt \ C_{ph,v}(t) .
\end{equation}
In these calculations, the starting configurations were taken from
the equilibrium configurations obtained with the parallel tempering
method and the subsequent time dependence was obtained from standard
MC simulations at each fixed temperature. The results shown on the
log-linear plot in Fig. 10 are indeed consistent with an activated
behavior of $\tau_{ph}$ and $\tau_v$. The straight lines in the plot
indicate that the data can be fitted to an  Arrhenius behavior, with
temperature independent energy barriers $E_{ph}=1.00$ and $
E_v=1.02$. In general, the energy barrier can be temperature
dependent, scaling with the correlation length \cite{fisher} as $E
\propto \xi^\Psi $. The observed Arrhenius behavior indicates that
these additional exponents are $\psi_{ph} \approx 0$, $\psi_{v}
\approx 0$. The behavior for the relaxation time $\tau_{ph}$ is
particularly important when the frustrated XY model is applied to
superconductors since it determines the temperature dependence of
the linear resistivity \cite{fisher,eg98,eg07}, $ \rho_L \propto
1/\tau_{ph}$. As a consequence, $\rho_L$ should be  finite at any
nonzero temperature but decrease exponentially as temperature
vanishes.

\begin{figure}
\includegraphics[bb= 1cm 3.5cm  19cm   16cm, width=7.5 cm]{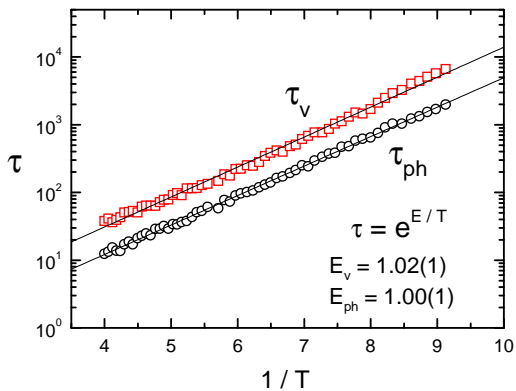}
\caption{ Temperature dependence of the relaxation times of phase
fluctuations $\tau_{ph}$ and vortex fluctuations $\tau_v$ for system
size $L=32$.}
\end{figure}

\section{Conclusions}
We have investigated the critical behavior of the fully frustrated
XY model on a two-dimensional honeycomb lattice by Monte Carlo
methods and a scaling analysis of the phase and vortex correlations.
No evidence of an equilibrium phase transition is found at nonzero
temperatures, in agreement with the conclusion of  ref.
\onlinecite{leeteitel} for a similar model. The absence of vortex
ordering at finite temperatures is also in agreement with the
estimates of ref. \onlinecite{korshu}, which find that this can only
be observed for very large system systems $(L > 10^5)$. However, our
finite-size scaling analysis is consistent with a zero-temperature
transition, where the critical temperature vanishes and phase and
vortex correlation lengths, $\xi_{ph}$ and $\xi_v$, diverge for
decreasing temperatures as a power law with a common critical
exponent, suggesting a coupled $T_c=0$ transition scenario. Since
both correlation lengths remain finite in the temperature range  $T
\ge 0.11$, where a  KT transition \cite{shih85} or a spin-glass
transition \cite{reid} was proposed to take place in earlier MC
simulations, our results also indicate that these apparent
transitions should be attributed to slow dynamics effects and not to
an equilibrium phase transition. Whether a vortex ordering
transition can, nevertheless,  occur for very large system sizes ($L
> 10^5$), as predicted in Ref. \onlinecite{korshu}, can not be
tested by our numerical simulation, which needs small system sizes
in order to insure full equilibration. The main features of the
correlation-length behavior obtained for the present model are the
same as those found in other frustrated XY models where the
phase-coherence temperature vanishes, such as, the two-dimensional
gauge glass model \cite{katz} and the XY model with irrational
frustration \cite{eg08}, but with different critical exponents. When
applied to superconductors, the divergent correlation length
$\xi_{ph}$ in these models determine both the linear an nonlinear
resistivity behavior, leading to a thermally activated linear
resistivity and nonlinear current-voltage scaling at low
temperatures \cite{fisher,eg98,eg07}. Similar behavior should be
observed for Josephson-junction arrays on a honeycomb lattice and
superconducting films with a triangular pattern of nanoholes
\cite{valles,triang} in a perpendicular magnetic field corresponding
to half flux quantum per plaquette, since both systems can be
modeled by a frustrated XY model on a honeycomb lattice. The
predicted thermal activated behavior for the linear resistivity
seems to be already been observed in the latter system
\cite{valles}.

\acknowledgements

The author wishes to thank  J. M. Valles Jr. for helpful discussions
and information on the experiments with superconducting nanohole
films, and C. S. O. Yokoi for helpful discussions on triangular
antiferromagnetic Ising models. This work was supported by Funda\c
c\~ao de Amparo \`a Pesquisa do Estado de S\~ao Paulo - FAPESP
(Grant 07/08492-9) and in part by computer facilities from Centro
Nacional de Processamento de Alto Desempenho em S\~ao Paulo -
CENAPAD-SP.


\begin{thebibliography}{00}
\bibitem{egunp} Additional simulations for the triangular
antiferromagnetic Ising model, confirmed that scaling of the
finite-size correlation length $\xi(L,T)$ is obtained when $x=
e^{-2/T} L$ is used as the scaling variable for $G(x)$, which is
consistent with the exponential behavior $\xi(T) \propto e^{2/T}$
obtained from the exact results of Ref. \onlinecite{stephenson}.
\bibitem{triang} The triangular lattice of nanholes in the ultrathin
films of Ref. \onlinecite{valles} can be regarded as the dual of the
honeycomb lattice where the frustrated XY model is defined.
\end{thebibliography}
\end{document}